\documentclass[preprint]{aastex}
\usepackage{graphicx}
\usepackage{amsmath,amssymb}
\usepackage{float}
\title{Prevailing dust-transport directions on comet 67P/Churyumov-Gerasimenko}
\author{Tobias~Kramer$^{a,b}$, Matthias~Noack$^{a}$\\
{\small $^a$Konrad-Zuse-Zentrum f\"ur Informationstechnik,  Takustr.~7, 14195 Berlin, Germany}\\
{\small $^b$Department of Physics, Harvard University, 12 Oxford St, Cambridge, MA 02138, U.S.A.}
}

\begin{document}

\begin{abstract}
Dust transport and deposition behind larger boulders on the comet 67P/Churyumov-Gerasimenko (67P/C-G) have been observed by the Rosetta mission. 
We present a mechanism for dust transport vectors based on a homogenous surface activity model incorporating in detail the topography of 67P/C-G.
The combination of gravitation, gas drag, and Coriolis force leads to specific dust transfer pathways, which for higher dust velocities fuel the near nucleus coma.
By distributing dust sources homogeneously across the whole cometary surface, we derive a global dust-transport map of 67P/C-G.
The transport vectors are in agreement with the reported wind-tail directions in the Philae descent area.
\end{abstract}

\maketitle

\section{Introduction}

Cometary nuclei display a wide variety of terrain and geological features, ranging from pits and mountains to flat or terraced regions. 
The ROLIS descent images of 67P/C-G show grains of regolith ranging from centimeter sizes to several meters, \cite{Mottola2015a}.
Behind these meter-sized obstacles wind tails are resolved in the ROLIS images.
Smaller sized particles as seen by ROLIS have been observed in orbit by the OSIRIS instrument, \cite{Sierks2015}, suggesting that dust particles contribute to both, the cometary tail originating from the nucleus and to the shaping of surface structures.
It has been proposed that the wind-tail structures are the result of boulders blocking impinging particles, which lead otherwise to abrasion of the surrounding terrain.
No model or explanation for the impinging-particle direction indicated by the wind tails has been given, besides the hypothesis that a particle source exists at some distance (\cite{Mottola2015a}).

Here, we establish a predictive model for the directions of impinging particles, which does away with a single particle-source and follows the opposite hypothesis: the whole cometary surface acts as source for dust grains.
The particular shape of the nucleus in combination with the Coriolis force leads then to prevailing transport directions in specific regions of the comet.
The same homogeneous activity model has been applied to dust particles lifted off the mantle with higher velocity and yields excellent agreement with the collimated dust structures seen by Rosetta around 67/C-G within a few cometary radii, \cite{Kramer2015a}.
The homogeneous surface-activity model works with a minimal set of parameters and assumptions, summarized as follows:
(i) incorporation of cometary topography on a scale down to $50$~m to ensure a dense and uniform sampling of the entire terrain, (ii) accurate computation of the gravitational potential for the partially concave shape of the nucleus, and (iii) full inclusion of the rotation of the nucleus.
In particular the rotation of the nucleus has a decisive role for establishing dust transport-vectors on the surface and for higher velocities into space.
Slower dust trajectories ($<1$~m/s) are strongly affected by the velocity dependent Coriolis force, which pushes particles on circular orbits within a few kilometers around the comet.

\section{Dust mobilization and transport}

In a coordinate-system attached to the cometary body, the acceleration of a dust particle is approximated by
\begin{eqnarray}\label{eq:accdust}
\vec{a}_{\text{dust}}(\vec{r})
&=&\vec{a}_\text{gas-drag}+\vec{a}_\text{grav}+\vec{a}_\text{centrifugal}+\vec{a}_\text{Coriolis}\\\nonumber
&=&\frac{1}{2}C_d \alpha N_\text{gas}  (\vec{r}) m_\text{gas} (\vec{v}_\text{gas}-\vec{v}_\text{dust})|\vec{v}_\text{gas}-\vec{v}_\text{dust}|
- \nabla \phi(\vec{r})-\vec{\omega}\times(\vec{\omega}\times\vec{r})-2\vec{\omega}\times \vec{v}_\text{dust},
\end{eqnarray}
which includes the acceleration of dust embedded in the more rapidly expanding gas, the gravitational force and the effect of rotation (centrifugal and Coriolis forces).
Additional forces, such as radiation pressure, changes in the rotation period or orientation of the rotation axis are neglected.
For instance, the acceleration due to radiation pressure on a 0.1~mm sized spherical dust particle at the perihelion (1.25~AU) is one order of magnitude smaller than the gravitational attraction by the nucleus.
We compute the gravitational potential of the irregular and partly concave shape of 67P/C-G by using a $39996$ triangles polyhedral representation.
Assumptions are a homogeneous density of the nucleus and a total mass of $10^{13}$~kg, \cite{Sierks2015}.
The polyhedral gravity equations given by \cite{Conway2014} are numerically efficiently implemented and the dust trajectories are integrated on parallel computing devices using a fourth order Runge-Kutta integration scheme.
In contrast to previous simulations done by \cite{Crifo2005a} we fully include the rotation of the comet.
In addition, the integration includes the gas-dust interaction, which has been neglected by \cite{Thomas2015}.
A stationary gas model around 67/C-G is constructed by assuming homogeneous outgassing activity across the whole cometary surface, \cite{Haser1957}, and thermal gas velocity (\cite{Huebner2006}, eq.~3.33),
\begin{equation}\label{eq:gasvel}
|\vec{v}_{\rm gas}|=\sqrt{\frac{k_B T}{m_\text{gas}}}.
\end{equation}
For CO$_2$ molecules at $T=200$~K (\cite{Ali-Lagoa2015a}) this yields $|\vec{v}_{\rm gas}|=200$~m/s. The CO$_2$ release by 67P/C-G has been detected by the ROSINA instrument on Rosetta at a heliocentric distance of 3.3 AU, \cite{Hassig2015a}.
Besides the velocity, the second parameter entering the Haser model is the surface number-density of the gas $N_\text{gas}(\vec{r}_\text{surface})$.
We neglect different outgassing rates that might originate from different illumination conditions on each surface element.
The transport mechanism described below relies on a slightly higher gravitational pull to the nucleus compared to the gas drag encountered by a dust particle around the comet.
The angular velocity vector and period are from \cite{Sierks2015}. 
\begin{figure}[t]
\begin{center}
\includegraphics[width=0.4\textwidth]{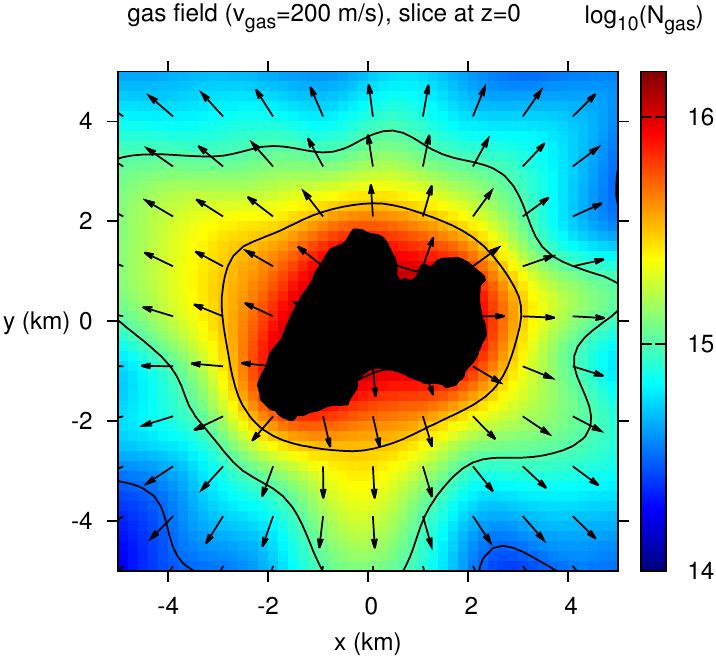}
\end{center}
\caption{\label{fig:gasfield}
Gas density and velocity direction across a slice at $z=0$~km through the cometary nucleus ($N_{\rm gas}=1.4 \times 10^{16}$~m$^{-3}$).
The highest gas density is reached in the concave neck region due to gas contributing from both side of the valley.
}
\end{figure}
The resulting gas number density distribution and velocity vectors are shown in Fig.~\ref{fig:gasfield}.
The maximal gas density is reached within the neck area of 67P/C-G due to the contribution of gas from the two sides of the neck valley, leading to an increased upward dust flow.
The gas drag depends on the dust-particle mass $m_\text{dust}$ and radius $R_\text{dust}$, as well as on the momentum of the impinging gas molecules.
For the standard value $C_d=2$ from \cite{Keller1994}
we obtain an overall factor $c_\text{gas-drag}$ for the gas-dust interaction
\begin{eqnarray}\label{eq:cgas}
\vec{a}_\text{gas-drag}
&=&c_\text{gas-drag}
\frac{(\vec{v}_\text{gas}-\vec{v}_\text{dust})|\vec{v}_\text{gas}-\vec{v}_\text{dust}|}{{|\vec{v}_\text{gas}|}^2},\\
c_\text{gas-drag}&=& k_B T N_\text{gas}(\vec{r}) \; \frac{\pi R_\text{dust}^2}{m_\text{dust}},\label{eq:pfgas}
\end{eqnarray}
where $R_\text{dust}, m_\text{dust}$ denote the particle radius and mass.
For re-colliding particles, the gas-drag forces pointing away from the comet are eventually compensated by the gravitational pull of the comet.
A re-collision is only possible if dust particles are emitted with a non-zero initial velocity acquired in the porous mantle layer of the comet, where gas-drag effects are amplified due to the compressed volume available for expansion, \cite{Huebner2006}.

We study dust transport occurring over an illuminated cometary surface with homogeneous gas emission.
Day/night effects of changing gas-density due to diurnal varying gas pressure are not considered. 
This limits the model to trajectories lasting less then half a cometary rotation period of about 6~h.
For the derived global dust transport map, we consider the complete comet as uniformly illuminated and dust emitting.
Hence, each trajectory represents a possible dust transport path over the sunlit part of the surface.
The irregular shape of the comet results in collisions with elevated terrain rotating into the pathway.
In addition, trajectories are affected by the modulations in the gravitational potential.

\section{Analysis of transport vectors}

The cometary surface is represented by a mesh of 39996 equal-sized triangles and each center of a triangle serves as origin for dust particles.
For the simulation, we chose an initial velocity as $0.1$~m/s along the triangle outwards normal to sample particles emitted with less than the escape velocity.
The results are independent of the number of triangles used to represent the comet shape and have been reproduced with the 19806 triangles mesh discussed in \cite{Kramer2015a}. 
The maximum gas particle density near the surface is varied from $1.1\times 10^{16}$~m$^{-3}$ to $1.4\times 10^{16}$~m$^{-3}$ for particles of density $1000$~kg/m$^3$ and radius $0.1$~mm, resulting in an overall gas drag very close to the gravitational force.
This procedure selects particles hovering up to several hours over the surface.
Note that only the overall prefactor $c_\text{gas-drag}\propto N_\text{gas}/R_\text{dust}$ is of importance for the gas drag, and thus, equivalently different particle sizes and gas densities are covered in the parametric study.
The results shown here for CO$_2$ hold also for a H$_2$O gas atmosphere at $T=266$~K with the $N_\text{gas}/R_\text{dust}$ ratio decreased by the factor $1.33$, see~eq.~(\ref{eq:pfgas}).

For a coarse-grained and ensemble-averaged picture, we embed the cometary shape in a volume mesh of cubes with side length $300$~m to track particles emitted from several surface triangles. 
In each cubic subvolume, we establish the average particle momentum direction and assign a total particle momentum of all traversing particles.
In addition, the number of particles and the velocities (direction and magnitude) are recorded.
The gas-drag forces delay the re-collision of the particle and lead to sustained motion above the surface while the comet rotates.
The gas forces are expected to temporally vary depending on the night/day exposure of the surface.
To focus on trajectories across always sunlit terrain,
we eliminate trajectories lasting longer than 6~h (half the rotation period of the comet).
We account for the expected downfall of particles caused by a reduction in up-ward gas drag, for instance by locally varying outgassing rates, self-shadowing, varying topographical features, and the day-night terminator
by taking into account all trajectories up to a ceiling of 300~m above the surface for the surface impinging particle flux.

\begin{figure}[t]
\includegraphics[width=0.35\textwidth]{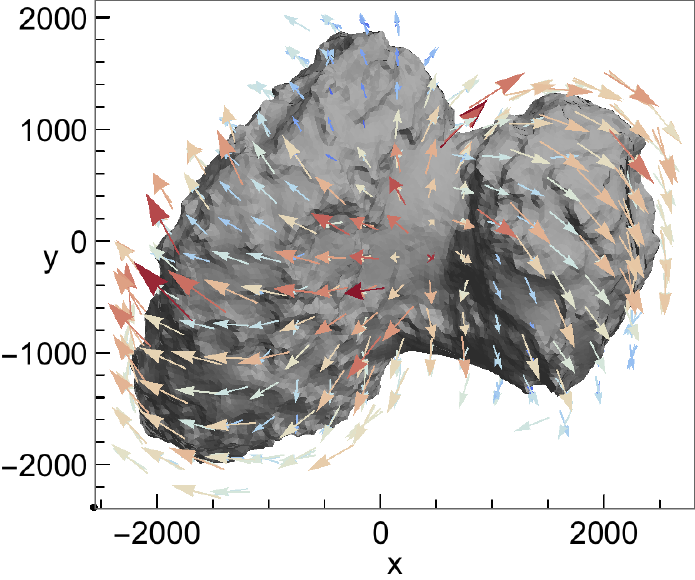}\hfill
\includegraphics[width=0.35\textwidth]{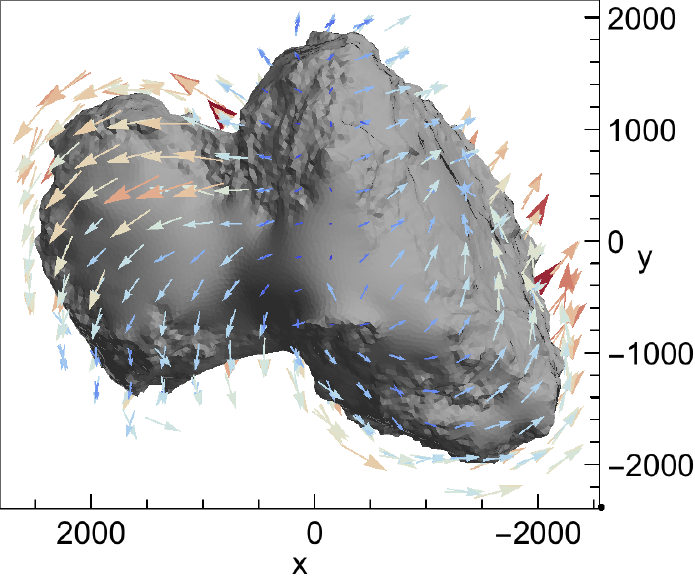}
\includegraphics[width=0.35\textwidth]{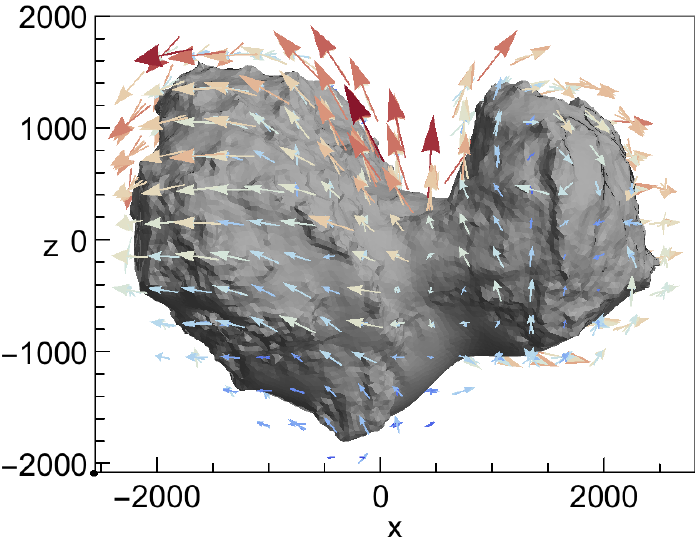}\hfill
\includegraphics[width=0.35\textwidth]{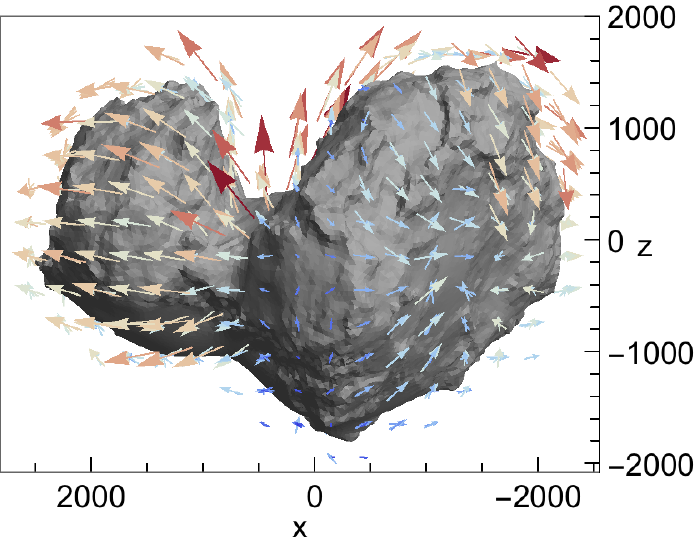}
\caption{\label{fig:velfield} 
Near surface vectors of dust transport around 67P/C-G. Arrows represent the momentum of the dust transport within the cubic subvolumes closest to the surface.
Color (blue to red) and arrow size denote increasing momentum per cube. 
Parameters: maximum gas particle density $N_{\rm gas}=1.4 \times 10^{16}$~m$^{-3}$, initial dust velocity $v_{i}=0.1$~m/s, homogeneous activity profile across the whole surface.
}
\end{figure}

\section{Discussion of prevailing dust transport directions}

The near surface vectors for the dust transport are shown in Fig.~\ref{fig:velfield}, where arrows represent the dust transport (product of velocity and mass) within the mesh cells closest to the surface.
The largest dust-transfer is predicted in the neck region, despite the homogeneous activity.
The reason is the increased outward gas drag above the concave neck due to contributory gas flow from both sides of the valley.
This process results in an increased upward acceleration and transfer of dust, despite the uniform surface distribution of initial dust grains.
The same effect is predicted in \cite{Kramer2015a} and observed for the dust jets forming above the neck region.

\begin{figure}[t]
\begin{tabular}{lll}
(a)&(b)&(c)\\
\includegraphics[width=0.26\textwidth]{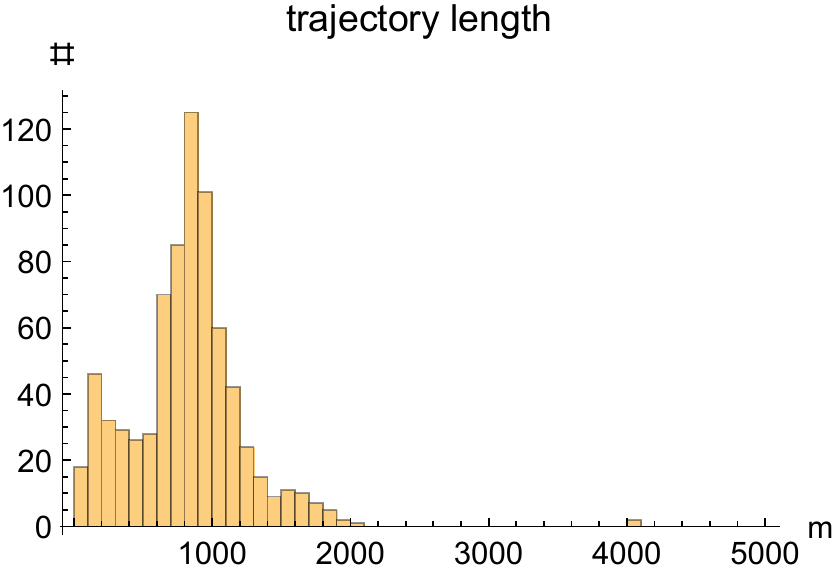}&
\includegraphics[width=0.25\textwidth]{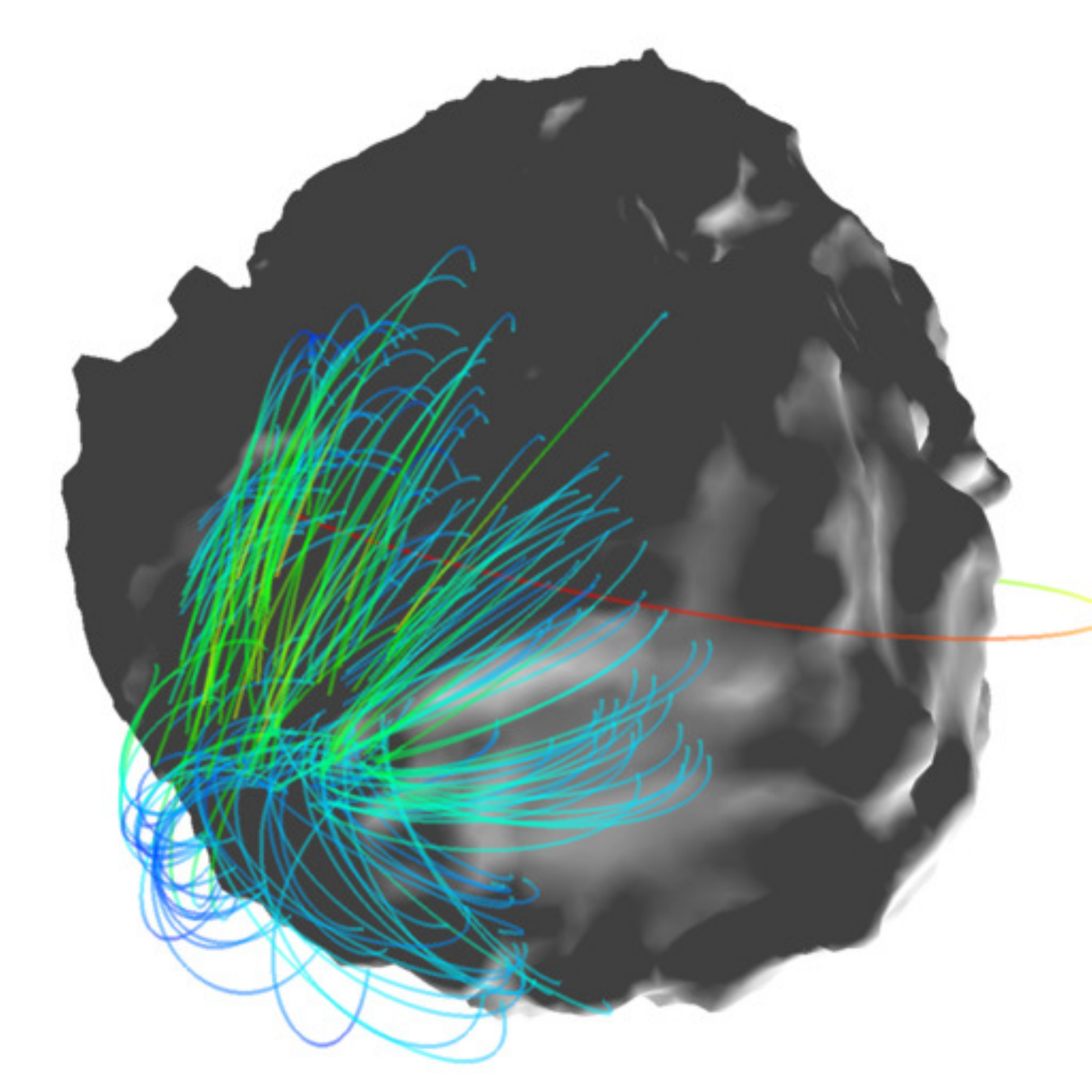}&
\includegraphics[width=0.25\textwidth]{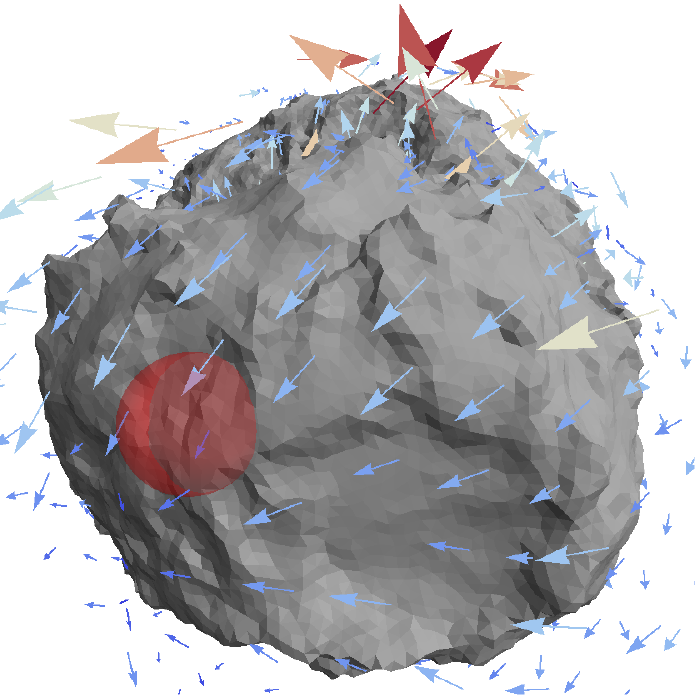}
\\
(d)&(e)&(f)\\\\
\includegraphics[width=0.26\textwidth]{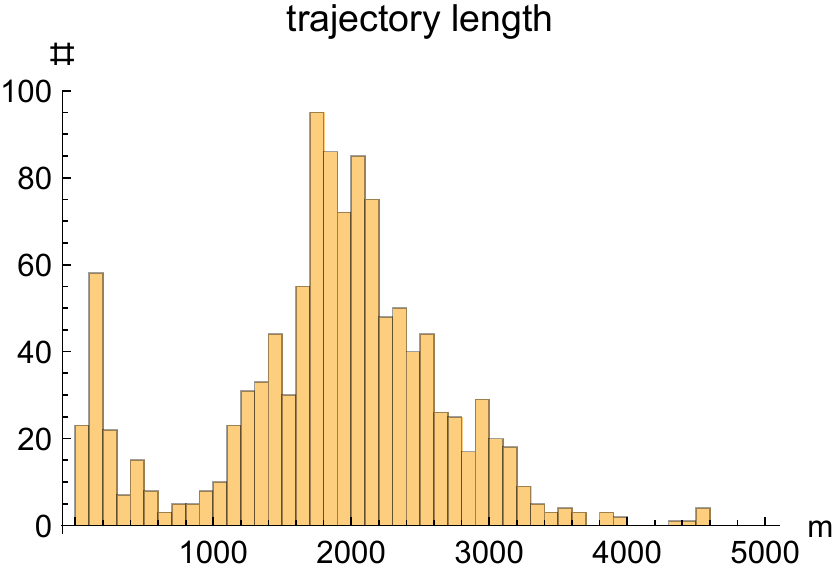}&
\includegraphics[width=0.25\textwidth]{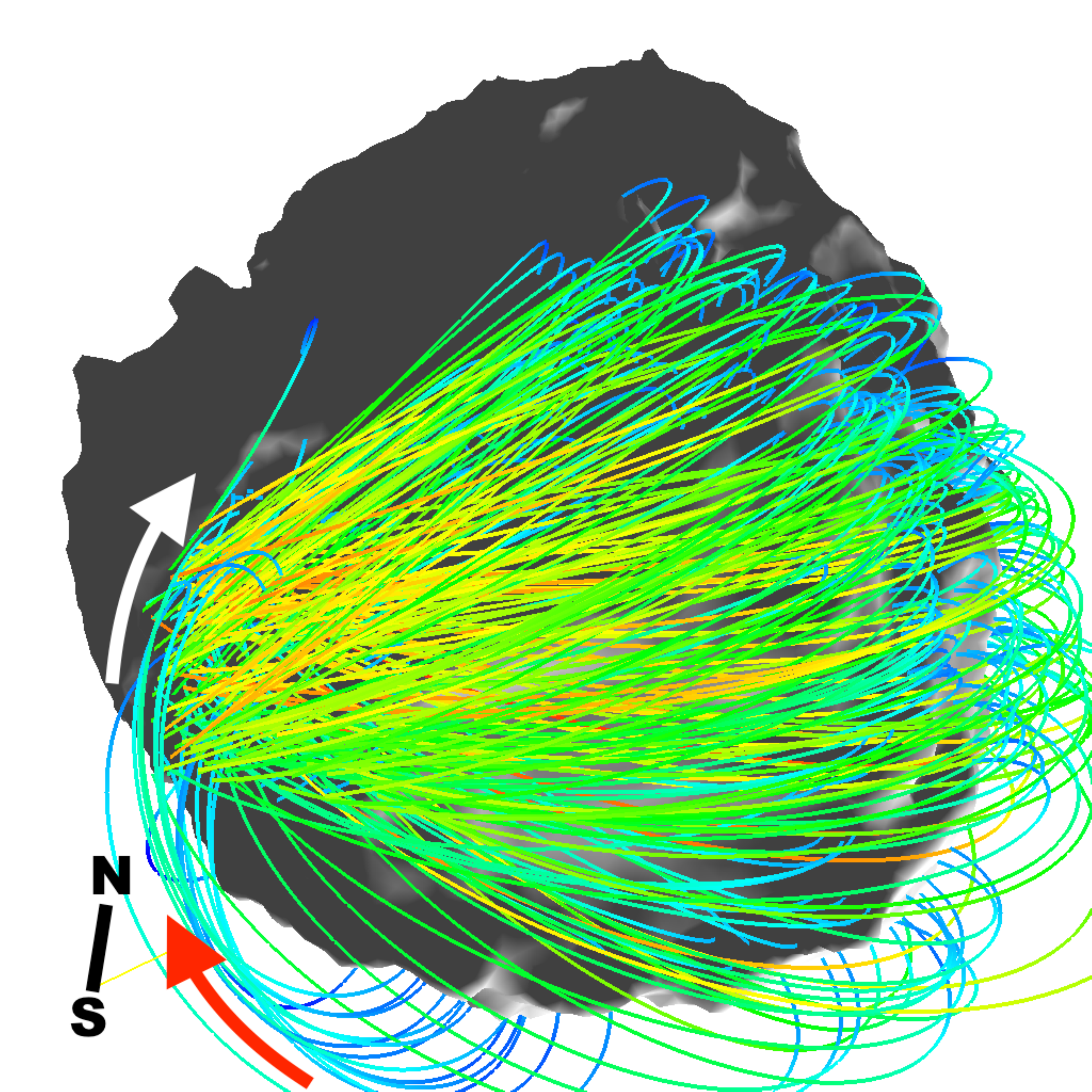}&
\includegraphics[width=0.25\textwidth]{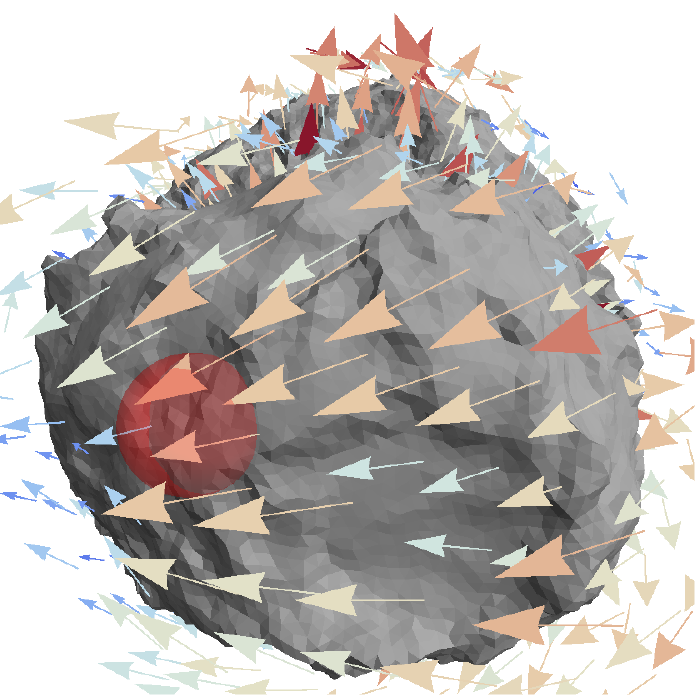}
\end{tabular}
\caption{\label{fig:velfieldROLIS}
Prevailing dust transport vectors over the small lobe of 67P/C-G for two different ratios of gas drag/gravity. 
Panels (a), (d) histogram of trajectory lengths. Panels (b), (e) sample of trajectories reaching a target sphere around the area imaged by Rosetta ROLIS, marked by the red sphere in (c), (f). Panels (c), (f) volume cell averaged transport vectors above the surface.
(a)-(c): gas drag for $N_\text{gas}=1.2\times 10^{16}$~m$^{-3}$ supports short-range transport (mean trajectory length $0.8$~km to the target region), while with increased gas drag (d)-(f), $N_\text{gas}=1.4\times 10^{16}$~m$^{-3}$, particles hover over the surface and bridge larger distances (mean trajectory length $1.7$~km to the target region). The dust trajectories marked by the red arrow in (e) coincide  with the observed wind tail direction (white arrow).
}
\end{figure}
Next, we investigate the dust transport vectors around the area imaged during the Philae/ROLIS descent, marked by the red target spheres in Fig.~\ref{fig:velfieldROLIS}~(c),~(f).
The viewpoint is chosen to approximately match the ROLIS image area, \cite{Mottola2015a} Suppl.~Mat.~Fig.~S1.
We select all trajectories entering the target region to identify the origins of impinging dust particles.
Typical trajectories are show in Fig.~\ref{fig:velfieldROLIS}~(b),~(e).
The dust trajectories start perpendicular to the local surface, but then move across the surface and possible around the lobe due to the cometary rotation.

Depending on how closely gas drag and gravitational force cancel each other, different travel distances and directions of the grains are predicted.
The distribution of trajectory lengths reaching the target area is given in Fig.~\ref{fig:velfieldROLIS}~(a),~(d).
The trajectories in Fig.~\ref{fig:velfieldROLIS}~(e) originate at a distance of on average $1.7$~km from the target area along two slightly inclined main transport vectors.
The dust stream indicated by the red arrow in Fig.~\ref{fig:velfieldROLIS}~(e) provides a possible source of impinging particles to explain the direction of the observed wind tail patterns by \cite{Mottola2015a} (white arrow).

\section{Conclusions}

We have proposed a mechanism for predicting dust migration on the surface of 67P/C-G based on homogeneous dust emission across the entire surface, linked to a detailed topographical model of the nucleus.
The homogeneous model yields a minimal-assumption prediction for the temporally-averaged dust transport.
Key aspects are the incorporation of the detailed topography, gravitational potential, and rotational forces of the comet, leading to global ``dust streams'' around 67P/C-G.
Previous approaches neglect the rotation of the nucleus due to the small ratio of centrifugal force to gravitational force and gas drag.
For dust transfer, the velocity dependent Coriolis effect dominates and rotation of the nucleus cannot be neglected. 
The predicted transport vectors are in line with the particle directions inferred from the ROLIS images.\\
The largest dust transport is expected away from the neck region, caused by an increased outward gas drag due to the focused transfer of gas momentum to dust by the concave valley shape.
This mechanism could act in addition to the increased thermal-stress hypothesis proposed by \cite{Ali-Lagoa2015a} in the neck region.\\
The same homogeneous activity-model has been applied to the near-nucleus coma by \cite{Kramer2015a} and is in excellent agreement with Rosetta NAVCAM observations of collimated dust jet/structures.
The present approach provides a global and unified theory of dust transfer on the surface and in the coma, in agreement with Rosetta observations.
Additionally, temporarily/spatially isolated sources of activity can be included in the model, but would require additional parameters acquired by observations.
Future observations by Rosetta are necessary to map out the global dust transport on 67P/C-G and compare it with predictions from theoretical models.

\paragraph{Acknowledgements}

We thank M.~Malmer for providing the initial shape model of comet 67P/C-G and the ESA/Rosetta NAVCAM team for releasing images of 67P/C-G on a continuous basis.
TK acknowledges support by a Heisenberg fellowship of the DFG grant KR 2889/5.
Computational resources and support by the North-German Supercomputing Alliance (HLRN) are gratefully acknowledged. 


\end{document}